\documentclass[3p,twocolumn,sort&compress]{elsarticle}

\usepackage{lineno,amsmath,amsfonts} %lineno,hyperref
\usepackage{tikz}
\usetikzlibrary{plotmarks}
\usetikzlibrary{arrows}
\usetikzlibrary[patterns]
\usetikzlibrary{shapes.geometric,arrows}
\modulolinenumbers[5]

\usepackage[draft]{changes} 
\definechangesauthor[color=blue]{FZ} 
\definechangesauthor[color=red]{NB}
\definechangesauthor[color=brown]{PP}

\usepackage{soul}
\soulregister\cite7
\soulregister\ref7
\soulregister\eqref7

\usepackage{hyperref}
\hypersetup{
    colorlinks=true,
    linkcolor=blue,
    filecolor=magenta,      
    urlcolor=cyan,
}

\newcommand{\rv}[1]{#1}

\newcommand{\vect}[1]{\boldsymbol{#1}}
\newcommand{\tensor}[1]{\boldsymbol{#1}}

\journal{the Journal of Computational Physics}

\begin{document}

\begin{frontmatter}

\title{Accelerating discrete dislocation dynamics simulations with graph neural networks}

\author[llnl]{Nicolas Bertin\corref{cor1}}
\ead{bertin1@llnl.gov}

\author[llnl]{Fei Zhou}

\cortext[cor1]{Corresponding author}
\address[llnl]{Lawrence Livermore National Laboratory, Livermore, CA, USA}

\begin{abstract}

Discrete dislocation dynamics (DDD) is a widely employed computational method to study plasticity at the mesoscale that connects the motion of dislocation lines to the macroscopic response of crystalline materials.
However, the computational cost of DDD simulations remains a bottleneck that limits its range of applicability.
Here, we introduce a new DDD-GNN framework in which the expensive time-integration of dislocation motion is entirely substituted by a graph neural network (GNN) model trained on DDD trajectories.
As a first application, we demonstrate the feasibility and potential of our method on a simple yet relevant model of a dislocation line gliding through \rv{an array} of obstacles.
We show that the DDD-GNN model is stable and reproduces very well unseen ground-truth DDD simulation responses for a range of straining rates and obstacle densities, without the need to explicitly compute nodal forces or dislocation mobilities during time-integration.
Our approach opens new promising avenues to accelerate DDD simulations and to incorporate more complex dislocation motion behaviors.

\end{abstract}

\end{frontmatter}

%\linenumbers

\section{Introduction}

Dislocations -- which are line defects in the crystal lattice -- define the plastic deformation of metals under most conditions.
Thus, discrete dislocation dynamics (DDD), which simulates the motion and interaction of an ensemble of dislocation lines, has become a very valuable computational tool for the study of metal plasticity at the mesoscale.
In DDD, dislocation lines are discretized into a series of segments connecting dislocation nodes, whose positions are evolved in time by integration of nodal velocities \cite{Cai06}.
Nodes motion results from the response of dislocations to the nodal driving force, which is a function of the local stress acting along the dislocation lines.
Since all dislocation lines elastically interact with each other, the calculation of nodal forces as required to time-integrate the system is typically the most expensive stage of the simulations.
The computational burden is further increased by the fact that time-integration in DDD simulations is made difficult by the intrinsic stiffness and highly non-linear behavior of the system, arising from the nature of dislocation interactions.
As a result, the timestep sizes used by traditional time-integrators are typically small, limiting the physical time-scale of the approach.
Thus, despite various recent progress and algorithmic advances \cite{bertin2020frontiers}, the computational cost of DDD simulations still remains a challenge that limits the range of applicability of the method.

Here we present a new DDD-GNN framework in which the computationally intensive time-integration procedure is fully replaced by a graph neural network (GNN) model trained to predict nodal displacements directly from the nodal configurations and applied loading.
The use of GNN appears particularly well suited for the task since dislocation line configurations are inherently a graph object.
Our model additionally takes advantage of the partitioning between short and long range elastic interactions commonly used in DDD.
By entirely bypassing explicit short-range forces calculation, our proposed framework has the potential to significantly accelerate the simulations.

Thus, the approach developed in this work differs from previous studies that applied machine-learning techniques to DDD, e.g. to accelerate the computation of individual interaction forces \cite{rafiei2020machine}, characterize dislocation microstructures \cite{steinberger2019machine}, or learn and predict crystal plasticity by focusing on the stress/strain curves \cite{salmenjoki2018machine, yang2020learning, minkowski2022machine}.

We demonstrate our approach by considering the fundamental case of a dislocation line gliding through \rv{an array} of obstacles under the action of an imposed loading.
This model constitutes an important scenario of dislocation dynamics, which has been widely employed to study dislocation-obstacle bypass of various classes \cite{foreman1966dislocation, scattergood1982strengthening, mohles2002computer, kulkarni2004effect, dong2010scaling, de2013dislocation, sobie2015analysis, santos2020multiscale, szajewski2021dislocation, cui2021statistical, szajewski2023statistical}, \rv{and is relevant for applications such as solid solution strengthening and precipitate and irradiation induced hardening}.
%and is characteristic of the plastic flow in single-slip conditions. 
Our DDD-GNN model is trained on ground-truth DDD data to predict nodal displacements (effective velocities) for a range of obstacle densities and straining rates.
We show that the trained GNN time-integrator model implemented within DDD is able to reproduce very well the local behavior of the dislocation motion, while accurately predicting the long-term simulation trajectories, such as the correct saturation flow stress under different loading regimes.

The remainder of the paper is organized as follows. In Section \ref{sec:method}, we start by briefly outlying the DDD method, after which we introduce our hybrid approach to incorporate the GNN time-integrator into the DDD framework.
In Section \ref{sec:forest}, we present the \rv{dislocation-obstacle glide} model used in this work to demonstrate our approach, and discuss the DDD data used to train our GNN model.
Results of our DDD-GNN model are then presented in Section \ref{sec:results} and the method is further discussed in Section \ref{sec:discussion}.
Finally, we conclude the paper in Section \ref{sec:conclusion}.

\section{DDD-GNN framework} \label{sec:method}

\subsection{DDD method} \label{sec:DDD}

\begin{figure}[t]
  \begin{center}
    \includegraphics[width=0.5\textwidth]{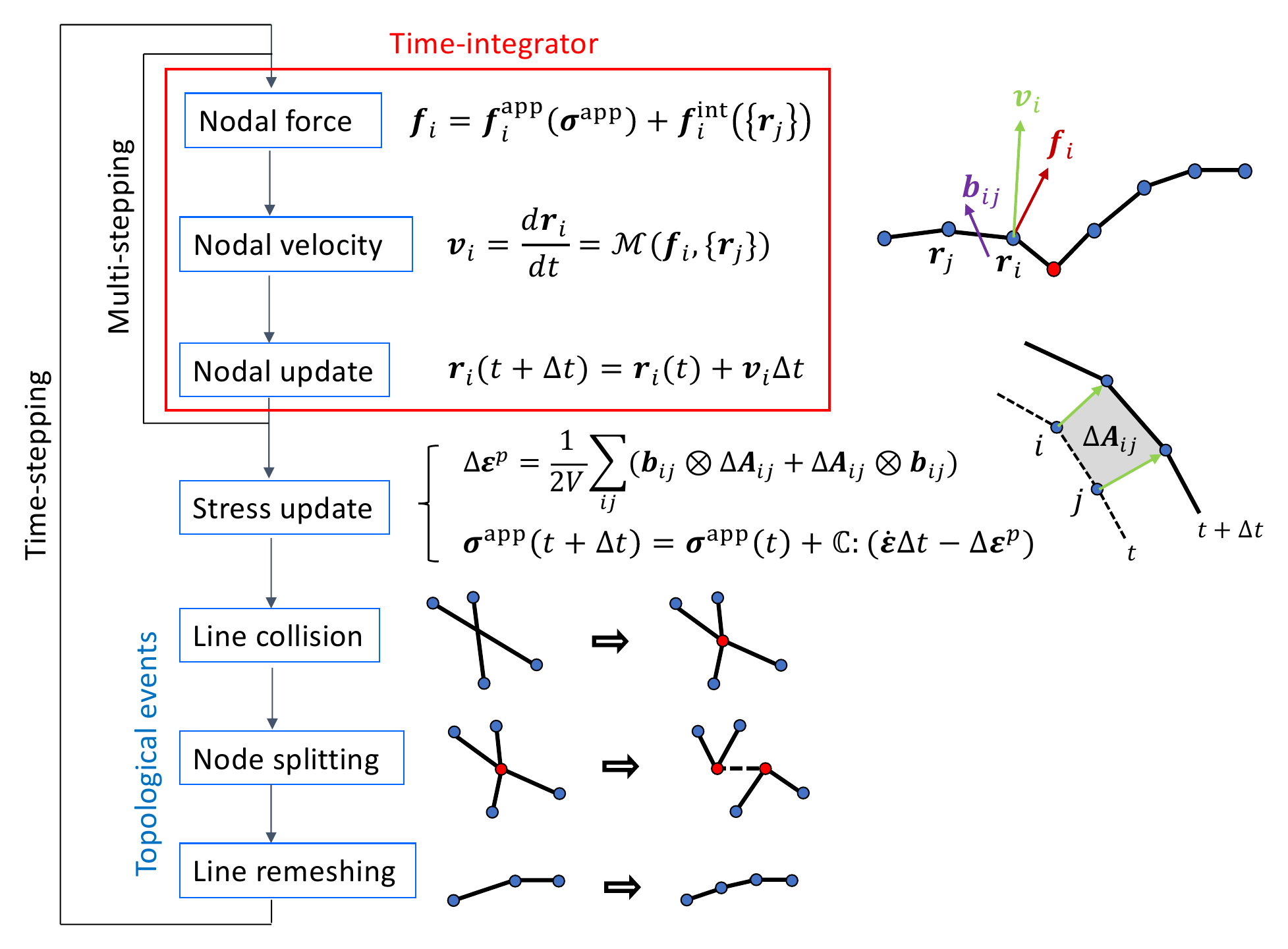}
  \end{center}
  %\vspace*{-0.5cm}
  \caption{Individual stages used to evolve the dislocation system at each time-step of a DDD simulation. In the DDD-GNN approach presented in this work we propose to fully replace the conventional time-integrator (red box) by a GNN model trained to predict nodal displacements directly from the dislocation configuration and applied loading.}
\label{fig:ddd_flowchart}
\end{figure}

In DDD, the dislocation configuration is represented as a graph $G = (\mathcal{V},\mathcal{E})$, where $\mathcal{V}=\{i\}$ is the collection of nodes (or vertices, used interchangeably in this work) with dislocation nodal positions $\vect{r}_i$, and $\vect{b}_{ij}$ is the Burgers vector of the dislocation segment connecting nodes $i$ to $j$, i.e.\ an edge $ ij \in \mathcal{E}$.
During a simulation, the system is evolved following the set of stages illustrated in the flowchart in Fig.~\ref{fig:ddd_flowchart}.
At each time-step, the nodal positions $\vect{r}_i$ are updated by time-integrating the equation of motion

\begin{equation} \label{eq:EOM}
\frac{d \vect{r}_i}{dt} = \mathcal{M}\left[ F(\left\{\vect{r}_j  \right\}) \right]
\end{equation}

\noindent where $\mathcal{M}$ is the mobility operator that describes the dislocation velocities in response to the forces $F$ exerted on the nodes \cite{Cai06}.
The force $\vect{f}_i \in F$ at each node $i$ is calculated by integrating the Peach-Koehler force along segments $ij$ connected to node $i$,

\begin{equation} \label{eq:Fi}
\vect{f}_i = \sum_j \int_{\vect{r}_i}^{\vect{r}_j} N(\vect{r}) \, \left( \tensor{\sigma}^{\rm tot}(\vect{r}) \cdot \vect{b}_{ij} \right) \times \frac{\vect{r}_j - \vect{r}_i}{\| \vect{r}_j - \vect{r}_i \|} \, d\vect{r}
\end{equation}

\noindent where $N$ is a shape function, and $\tensor{\sigma}^{\rm tot}$ denotes the total stress acting along the dislocation line \cite{Arsenlis07}. The total stress tensor generally includes two main contributions, $\tensor{\sigma}^{\rm tot} = \tensor{\sigma}^{\rm app} + \tensor{\sigma}^{\rm int}$, where $\tensor{\sigma}^{\rm app}$ is a uniform applied stress resulting from the imposed loading, and $\tensor{\sigma}^{\rm int}$ is the internal stress contribution from all dislocation segments in the medium.
Consequently, the nodal force in Eq.~\eqref{eq:Fi} can be written as

\begin{equation} \label{eq:Fi2}
\vect{f}_i = \vect{f}_i^{\rm app}(\tensor{\sigma}^{\rm app}) + \vect{f}_i^{\rm int}
\end{equation}

\noindent where $\vect{f}_i^{\rm app}$ denotes the applied loading force, and $\vect{f}_i^{\rm int}$ is the internal force contribution, the calculation of which depends on all segments pair interactions (and their periodic images when using periodic boundary conditions), and whose brute-force computation scales as $\mathcal{O}(|\mathcal{E}|^2)$.

To alleviate the computational cost of the method, several approaches that take advantage of the $1/R$ decay of dislocation stress fields have been devised.
The general idea is to limit the explicit computation of segment pair forces to close, short-range interactions, while approximating longer range interactions using various techniques. These include the fast multipole method (FMM) \cite{LeSar02, Arsenlis07}, finite element method (FEM) \cite{Lemarchand01, Vattre14, jamond2016consistent}, or fast Fourier transforms (FFT) approaches \cite{Bertin15, Bertin18a, bertin2019connecting}.

Following this partitioning of the stress, the internal contribution of the nodal force in Eq.~\eqref{eq:Fi2} can therefore be decomposed as,

\begin{equation} \label{eq:Fi3}
\vect{f}_i^{\rm int} = \sum_{kl \in \mathcal{E}} \vect{f}_i^{kl} \simeq \underbrace{\sum_{kl \in \mathcal{E} < r_c} \vect{f}_i^{kl}}_{\vect{f}_i^{\rm short}} + \underbrace{\vect{f}_i^{\rm long}(\tensor{\sigma}^{\rm long})}_{\vect{f}_i^{\rm long}}
\end{equation}

\noindent where $\vect{f}_i^{kl}$ denotes the explicit force contribution of segment $kl$ to node $i$, and the notation $kl \in \mathcal{E} < r_c$ designates all segments $kl$ of the graph that are closer to node $i$ than the short-range cut-off distance $r_c$.
Thus, $\vect{f}_i^{\rm long}$ is the long-range dislocation interaction contribution, e.g. approximated from a stress contribution $\tensor{\sigma}^{\rm long}$ computed with one of the techniques mentioned previously.
$\vect{f}_i^{\rm short}$ is the short-range contribution, whose calculation involves explicit evaluation of elastic interactions between neighboring segment pairs.
In practice, the computation of this last contribution, $\vect{f}_i^{\rm short}$, remains the most expensive operation in DDD simulations, which can account for more than 90\% of the total computation time \cite{bertin2019gpu}.

Once nodal forces are calculated, the set of nodal velocities $\vect{v}_i$ are computed and the equation of motion \eqref{eq:EOM} is integrated during which the nodal positions $\{\vect{r}_i\}$ are advanced in time.
The time-step size is typically selected based on an error tolerance scheme \cite{Arsenlis07}, or by employing more sophisticated techniques such as multi-stepping or subcycling algorithms \cite{Sills14, Sills16}.

After time-integration is performed, the area swept by all dislocations during the time interval $\Delta t$ is computed to evaluate the increment of plastic strain $\Delta \tensor{\epsilon}^p$ generated by dislocations motion

\begin{equation}
\Delta \tensor{\epsilon}^p = \frac{1}{2V} \sum_{ij} \left( \vect{b}_{ij} \otimes \Delta \vect{A}_{ij} + \Delta \vect{A}_{ij} \otimes \vect{b}_{ij} \right)
\end{equation}

\noindent where $V$ is the simulation volume, and $\Delta \vect{A}_{ij}$ is the vector of magnitude equal to the area swept by segment $ij$ over time interval and of direction normal to the plane in which segment motion has occurred.

When the system is loaded under an imposed strain rate $\dot{\tensor{\epsilon}}$, the applied stress is further incremented by a value of

\begin{equation}
\Delta \tensor{\sigma}^{\rm app} = \mathbb{C} : \left( \dot{\tensor{\epsilon}} \Delta t - \Delta \tensor{\epsilon}^p \right)
\end{equation}

\noindent where $\mathbb{C}$ is the fourth-order tensor of elastic moduli.

Finally, topological operations are performed to handle dislocation core reactions, e.g. dislocation collisions and junction nodes splitting, and remeshing of the dislocation lines, see Fig.~\ref{fig:ddd_flowchart}. For a more comprehensive overview of the DDD method the reader is referred to Refs.~\cite{Cai06, Arsenlis07}.

\subsection{Machine-learned time-integrator} \label{sec:DDDGNN}

\begin{figure*}[t]
  \begin{center}
    \includegraphics[width=0.95\textwidth]{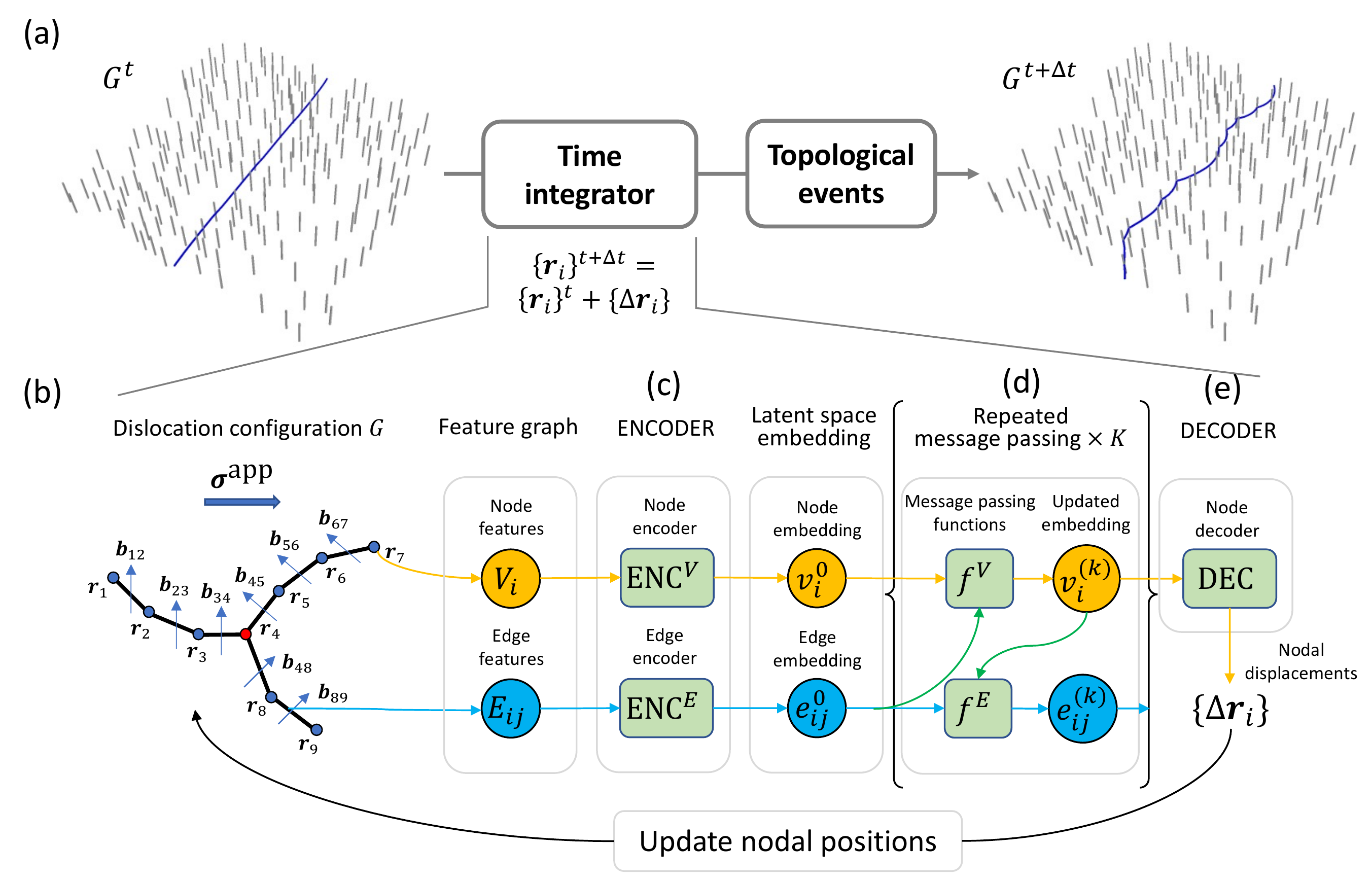}
  \end{center}
  %\vspace*{-0.5cm}
  \caption{(a) Our hybrid DDD-GNN approach aims at predicting the state of the dislocation configuration at the next time increment by replacing the conventional time-integration procedure with a GNN model while leaving the treatment of topological events to the traditional DDD framework. (b) GNN architecture mapping dislocation network $G$ to nodal displacements in Eq.~\eqref{eq:whole-map}. (c) Node and edge encoders that translate input node variables (node type and applied stress $\tensor{\sigma}^{\rm app}$) and edge variables (Burgers vector, line segment vector, and long-range stress $\tensor{\sigma}^{\rm long}$) into latent space representations, respectively. (d) Stacked message-passing layers that gather latent nodal and edge variables on each edge as updated edge representations and then pass and aggregate them to connected nodes. (e) Node decoder for the regression task of predicting nodal displacements $\{\Delta\vect{r}_i\}$.}
\label{fig:ddd_gnn}
\end{figure*}

Following the description in Section \ref{sec:DDD}, the time integration procedure is central to the DDD approach and typically involves a large number of forces and velocities calculations for all nodes. Here we view instead the time evolution of the dislocation configuration as a machine-learning task defined on a graph network.
To handle this task, we propose an hybrid DDD approach whereby a GNN model is used to replace the time-integration stage, Fig.~\ref{fig:ddd_gnn}, while the rest of the DDD cycle (e.g. topological operations) is handled using the traditional DDD treatment.

We introduce a GNN-based time-integrator as a general learnable model $\mathcal{H}: G^t \rightarrow G^{t+\Delta t}$ trained to predict the state of the dislocation graph configuration at the next instant in time $t+\Delta t$ given the state $G^t$ at the current time $t$.
Since the topology of the dislocation network does not change during the time-integration procedure, the configuration $G^{t+\Delta t}$ can be predicted by learning the nodal displacements $\Delta \vect{r}_i$ resulting from integration of the equation of motion~\eqref{eq:EOM} over the given time interval $\Delta t$.
The learnable model is therefore

\begin{equation} \label{eq:whole-map}
    \mathcal{H}(G) = \{\Delta \vect{r}_i\}.
\end{equation}

Eq.~\eqref{eq:whole-map} is essentially a map from graph to vector per node, which is computationally similar to the task of constructing force fields or interatomic potentials, a very active research field of materials science \cite{Deringer2019}.
Such interatomic potentials, especially when trained from quantum mechanical calculations, allow for orders of magnitude speed-ups in molecular dynamics (MD) simulations compared to \textit{ab initio} MD with similar accuracy, e.g. \cite{GAP}.
Tremendous progress has been made utilizing either traditional force field models based on physical intuition of the underlying materials or ML methods that often take inspirations from rapid developments in deep neural networks. While the former often offer better efficiency, interpretability and generalizability, they may be more time-consuming to develop, especially when applied to complicated structures with a lack of quantitative physical justifications \cite{Senftle2016, Frenkel-Molecular-Simulation, leach2001molecular}. Modern ML-based models, on the other hand, might be less efficient, interpretable or generalizable, but could be trained with enough data even in the absence of deep physical knowledge \cite{Deringer2019, Mueller2020, Noe2020, Unke2021, Behler2021}. Given our goal of mapping a relatively complicated target property -- nodal displacements integrated over multiple steps -- we take the latter approach.

One critical aspect for the success of the model lies in the partitioning of stresses (equivalently forces) between short-range and long-range contributions, as discussed in Section \ref{sec:DDD}. Here we capitalize on the fact that short-range interactions are fully characterized by the local dislocation configuration, while the long-range interactions are by definition a function of the full dislocation network.
In our model we take advantage of this partitioning and let the GNN learn the short-range interactions, while providing the long-range stress information as an input of the dislocation configurations.

\subsection{Message-passing Graph Neural Network} \label{sec:GNN}

We approximate the local dependence of short-range interactions on the neighboring dislocation configuration with a message-passing GNN \cite{battaglia2018relational}, one of the most promising ML approaches with increasing number of successful applications for predicting force fields and other materials properties \cite{Xie2018, Chen2018, Park2021, Batzner2021} and simulating complex physics \cite{sanchez2020learning}.
\rv{As shown at the beginning of Section \ref{sec:DDD}, a dislocation configuration can be represented by a graph $G = (\mathcal{V},\mathcal{E})$, where $\mathcal{V} = \{V_i\}$ is a collection of dislocation node/vertex features (attributes), and $\mathcal{E} = \{E_{ij}\}$ is a collection of dislocation segment/edge features.}
A GNN is an algorithm that naturally operates on such graph-structured data.
Specifically, let the input features for each node $i$ be 
\begin{align} \label{eq:node-feature}
V_i = (t_i,  \tensor{\sigma}^{\rm app}),
\end{align} 
where $t_i$ is a flag that can be used to specify the type of node $i$ (e.g. pinned or junction node).
The uniform input stress $\tensor{\sigma}^{\rm app}$ can be incorporated into the graph as either an edge or vertex feature, and in this work was assigned  to $V_i$.
The input edge features on edge $ij$ are 
\begin{align} \label{eq:edge-feature}
E_{ij} =( \vect{b}_{ij}, \vect{r}_{j}-\vect{r}_{i}, \tensor{\sigma}^{\rm long}_{ij}),
\end{align}
\rv{where $\vect{r}_{j}-\vect{r}_{i}$ are the local segments line vectors, naturally compatible with the use of periodic boundary conditions.}
To satisfy Burgers vector conservation, we assume a directed graph, i.e. if $ij$ is an edge with Burgers vector $\vect{b}_{ij}$, then $ji$ is also an edge but with opposite Burgers vector $-\vect{b}_{ij}$.
\rv{The vertex and edge features in Eqs.~(\ref{eq:node-feature}-\ref{eq:edge-feature}) contain all information of the dislocation network necessary to predict the nodal displacements at the next instant in time. To do so, our GNN model follows \cite{MeshGraphNet} and} is first composed of vertex and edge encoders $\text{ENC}^V$, $\text{ENC}^E$ transforming concatenated input features into a latent space (see Fig.~\ref{fig:ddd_gnn}c):
\begin{align} \label{eq:encoder}
    v^{(0)}_i = \text{ENC}^V(V_i), \ e^{(0)}_{ij} = \text{ENC}^E(E_{ij}),
\end{align}
followed by $K$ stacked message passing layers $f^{E(k)}$, $f^{V(k)}$ ($1 \leq k \leq K$) operating on the latent vertex and edge variables (Fig.~\ref{fig:ddd_gnn}d):
\begin{align}
e^{(k)}_{ij} &= f^{E(k)}( e^{(k-1)}_{ij}, v^{(k-1)}_i, v^{(k-1)}_j ),  \\
v^{(k)}_{i} &= f^{V(k)}( v^{(k-1)}_i, \sum_j e^{(k)}_{ij}), \label{eq:MP}
\end{align}
and finally a node decoder $\text{DEC}$ that translates the latent node variables $v^{(K)}$ into the desired properties (Fig.~\ref{fig:ddd_gnn}e), i.e. nodes displacement over the prescribed time integration steps:
\begin{align} \label{eq:decoder}
    \Delta\vect{r}_i  = \text{DEC}(v^{(K)}_i).
\end{align}
Together, Eqs.~(\ref{eq:node-feature}-\ref{eq:decoder}) constitute a GNN implementation of the surrogate model in Eq.~\eqref{eq:whole-map}.
\rv{In short, the trainable model $\mathcal{H}(\{V_i\},\{E_{ij}\}) = \{\Delta \vect{r}_i\}$ takes the collection of dislocation node and segment features defined in Eqs.~(\ref{eq:node-feature}-\ref{eq:edge-feature}) as inputs and predicts the nodal displacements $\{\Delta \vect{r}_i\}$ as outputs. During training, the internal parameters of the neural network operators (functions $\text{ENC}^V$, $\text{ENC}^E$, $f^V$, $f^E$, and $\text{DEC}$) are adjusted so as to learn to approximate the relation between the local graph structure and the resulting nodal displacements, thereby bypassing the need for explicit short-range force calculations.}
The critical message-passing steps in Eq.~\eqref{eq:MP} allow the GNN to transmit and aggregate information along connected edges, and each additional message-passing layer builds up a more complex dependence of the desired properties on vertices farther away.
With this approach, the number of message-passing layers $K$ required to obtain accurate predictions is then directly related to the cut-off distance used to distinguish between short-range and long-range interactions, e.g. as set by the FMM cell structure, or the FFT grid resolution.
Our approach still requires the explicit computation of long-range stresses integrated on the segments $\tensor{\sigma}^{\rm long}_{ij}$ as done in the traditional DDD method. However this operation is typically relatively cheap compared to the explicit short-range interaction calculation. \rv{(The long-range operation might in principle be replicated by many GNN message-passing layers, but doing so would be computationally cumbersome and inefficient.)}

Note that in the present work, we simply adopted the dislocation line segments as the edges of the GNN, as this simple approach turns out to be sufficient for our first application presented in Section \ref{sec:forest}. However, the GNN approach itself is not limited to such simplifications, and additional edges may be incorporated to capture more complex interactions between nodes that are not directly connected by dislocation line segments for more general applications, e.g.\ in future works on bulk dislocation dynamics.
We also note that in contrast to other ML techniques such as convolutional neural networks (CNN) that typically require inputs of fixed data structures, once trained GNN models based on message-passing can be employed to make predictions for input graphs of arbitrary size and connectivity.
This property is essential in the context of DDD simulations in which the number of dislocations nodes and segments constantly evolves during the simulated trajectories as a result of topological events and dislocation multiplication.

We use an in-house GNN implementation in PyTorch \cite{NEURIPS2019_9015} to build and train the model as a regression task. \rv{All neural network operators are built from multi-layer perceptrons with two hidden layers, layer normalization \cite{LayerNorm} between them, and skip connections \cite{resnet} (only in the message-passing layers). Instead of the ReLU activation function used in Ref.~\cite{MeshGraphNet}, the GELU activation function \cite{GELU} was found to consistently outperform and therefore adopted throughout.
The trained time-integrator is then directly implemented within the ParaDiS DDD program \cite{Arsenlis07} using the C++ PyTorch interface.} All neural network computations are performed with single (32-bit) floating-point precision.

\section{Application: the \rv{dislocation-obstacle glide} model} \label{sec:forest}

\subsection{DDD simulations setting}

\begin{figure}[t]
  \begin{center}
    \includegraphics[scale=0.6]{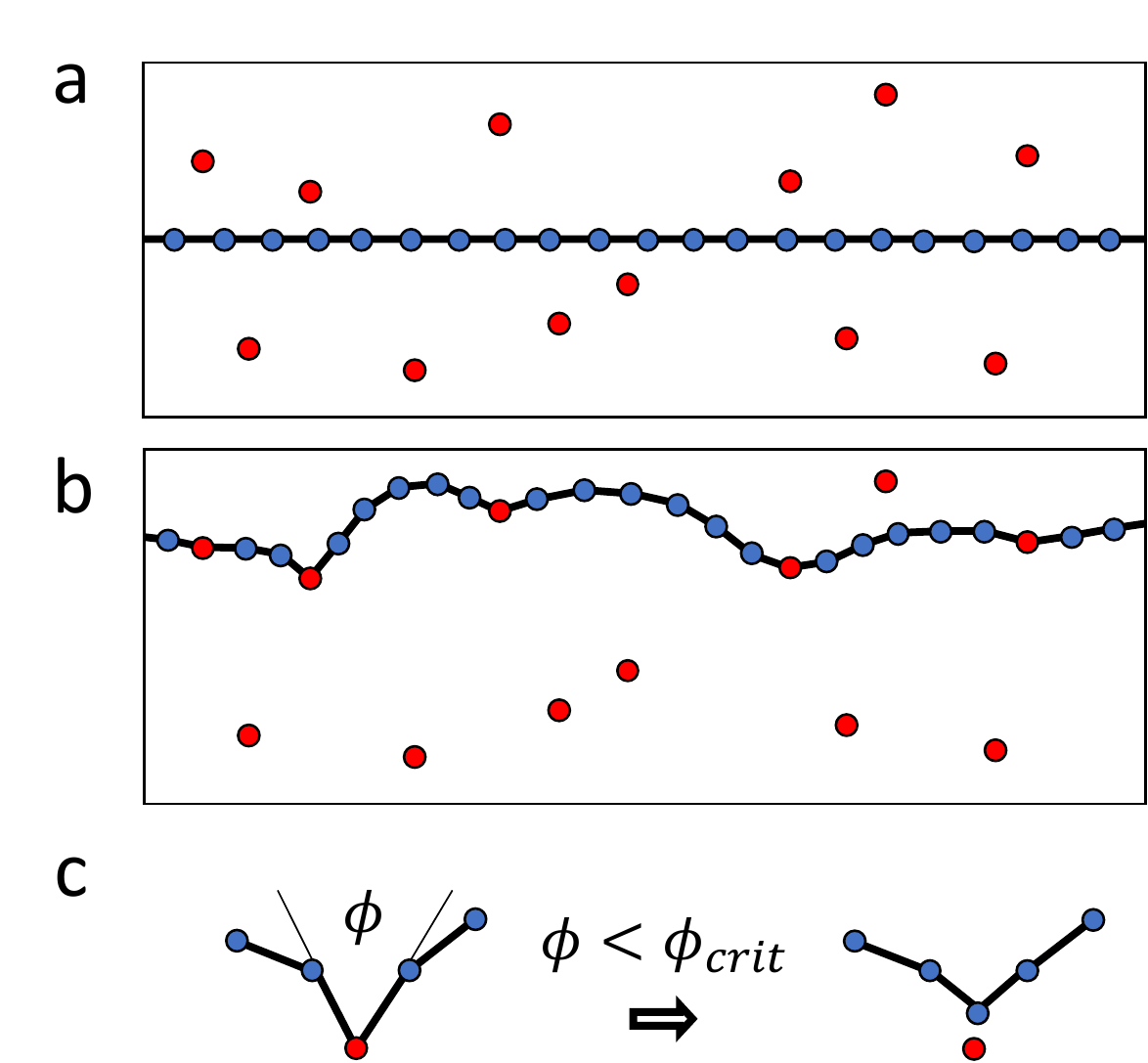}
  \end{center}
  %\vspace*{-0.5cm}
  \caption{Schematic of the \rv{dislocation-obstacle glide} model employed in this work. (a) An initial infinite discrete dislocation line (black line) is introduced within the central glide plane of a simulation box. A random array of obstacles are also introduced in the dislocation glide plane (red points). (b) Upon loading the dislocation line moves and comes in contact with the obstacles. Dislocation nodes in contact with obstacles are pinned. (c) A breakaway angle criterion is used to unpin the dislocation from the obstacles. The pinning node is released from the obstacle when its two connected arms form an angle $\phi$ smaller than a critical breakaway angle $\phi_{\rm crit}$.}
\label{fig:forest}
\end{figure}

To demonstrate the feasibility and validity of our approach, we consider in this work a model of a dislocation line gliding through \rv{an array} of obstacles.
The \rv{dislocation-obstacles} simulation setting is illustrated in Fig.~\ref{fig:forest}. An infinite dislocation line of edge character is introduced on the central glide plane of a simulation box.
A random array of point-like obstacles is further initialized on the glide plane of the dislocation.
The dislocation line is then set in motion by application of a constant strain rate $\dot{\epsilon}$ along a uniaxial loading direction inclined at $45^o$ from the glide plane in the direction of the dislocation Burgers vector.

Upon contact of the dislocation with an obstacle, the dislocation is pinned at the obstacle position, with local velocity set to zero at the pinning node.
Following \cite{foreman1966dislocation}, a breakaway angle criterion is then used to determine when the dislocation gets unpinned from the obstacle.
As illustrated in Fig.~\ref{fig:forest}c, when the two segments connected to a pinning node form an angle $\phi < \phi_{\rm crit}$ smaller than a critical breakaway angle $\phi_{\rm crit}$, the dislocation pinning node is released from the obstacle and its velocity is no longer set to zero.

During the simulations cross-slip is disabled so that the moving dislocation line remains on its glide plane.
Periodic boundary conditions are applied along the directions of the dislocation glide plane.

\subsection{DDD generated training dataset}

To train our GNN model we first generate ground-truth DDD trajectories in which dislocation graph configurations, applied and long-range stresses, and nodal displacements are recorded at a frequent interval of time during the simulations. 
For each node in the configurations nodal flag $t_i$ (Eq.~\eqref{eq:node-feature}) is set to 0 or 1 depending on whether node $i$ is pinned at an obstacle or not.
A total of 100 simulations is run for 10,000 time-steps each and configurations are recorded every 100 steps, i.e. generating a total of 10,000 training configurations.

For each training simulation we use a volume $V = L \times L \times H$, with $L^2 = (10 {\rm \mu m})^2$ is the area of the dislocation glide plane, and $H = 2 {\rm \mu m}$ is the dimension of the box along the glide plane normal.
The obstacles density $\rho_{\rm obs}$ in the glide plane is chosen at random between values of $1 \times 10^{12}$ and $3 \times 10^{12}$ 1/m$^2$ (corresponding to a number of $100$ to $300$ obstacles on the glide plane), while the strain rate $\dot{\epsilon}$ is chosen in the range $3 \times 10^2$ to $3 \times 10^3$ 1/s.
The breakaway angle is set to $\phi_{\rm crit} = 90^o$, \rv{consistent with the bypass of a small precipitate at finite temperature \cite{bacon2009dislocation}.}
We use an isotropic, linear mobility law with drag coefficient $B = 10^{-4}$ Pa/s. The Burgers vector magnitude is set to $b = 0.25$ nm, and we use a shear modulus of $\mu = 50$ GPa and Poisson's ratio of $\nu = 0.3$.

Time-integration is performed with a multi-stepping time integrator with fixed global time-step size of $1 \times 10^{-10}$s. In this scheme, a Heun (trapezoidal) base integrator is used to advance nodal positions with adaptive step size selected based on an error tolerance of $r_{\rm tol} = 1b$. Multi-stepping is then achieved by sequentially repeating this operation until the global time-step size is reached. We find that a total of 10 to 20 substeps is typically required to time-integrate the system for a single time-step for our choice of the global time-step size.
Long-range stress contributions are computed using the FMM method with $4^3$ cells.

Note that for this choice of the simulation parameters, the drag stress (defined as the stress required to move the dislocation in the absence of any obstacle) given by $\tau^{\rm drag} = 2 \dot{\epsilon}LHB/b^2$ ranges from 19.2 to 192 MPa, which is generally higher than the obstacles stress which is on the order of $\tau^{\rm obs} \simeq \mu b\sqrt{\rho_{\rm obs}}$ and whose value ranges from 12.5 to 21.6 MPa.
Thus the ground-truth DDD data generated to train the GNN model corresponds to simulations performed in the drag-controlled regime $\tau^{\rm drag} > \tau^{\rm obs}$, where obstacles are not expected to play a significant role in the determination of the crystal's flow stress. As will become clear later, the choice of such a regime for the training process is meant to illustrate the robustness of the approach.

\begin{figure}[t]

\begin{center}
  (a)\includegraphics[scale=0.3]{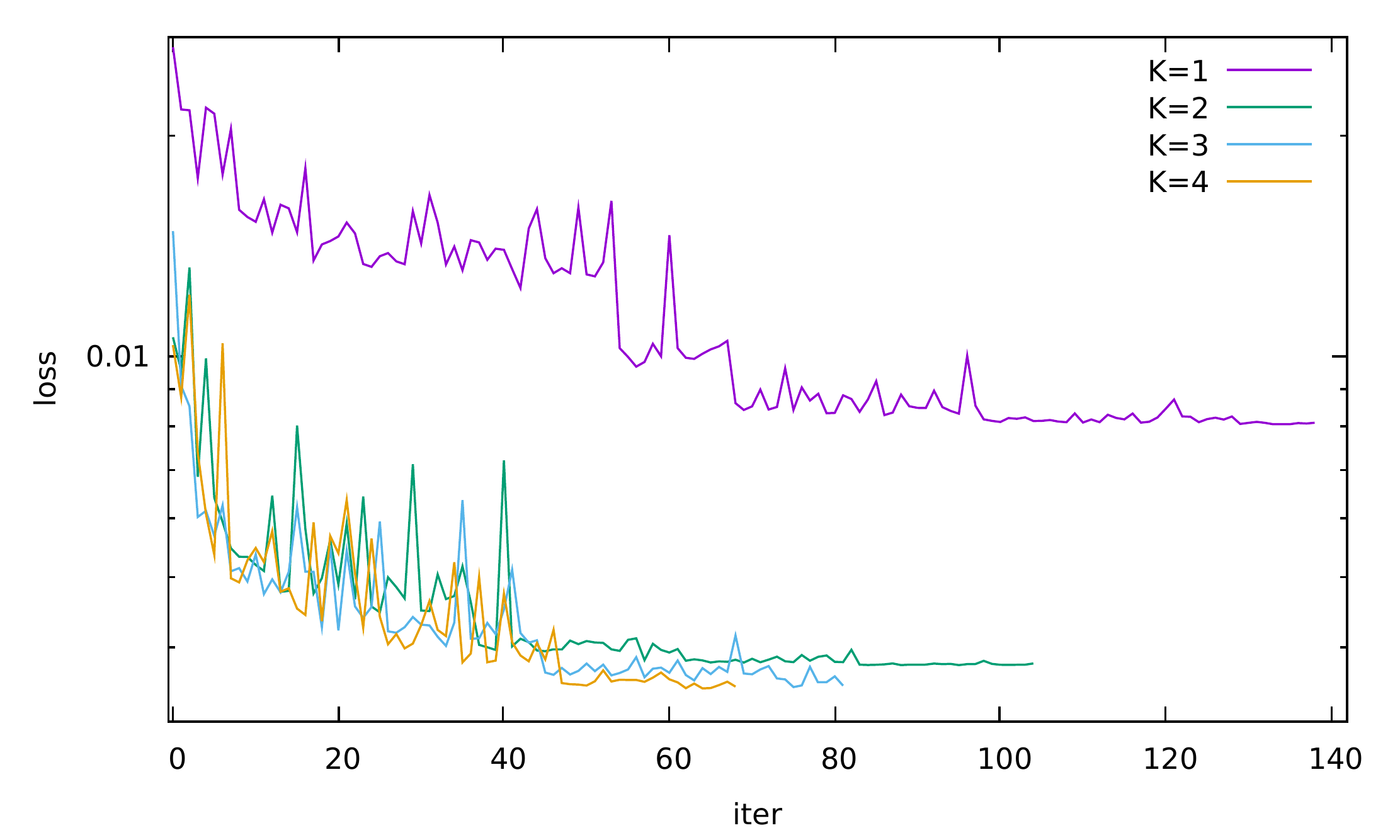}
  (b)\includegraphics[scale=0.5]{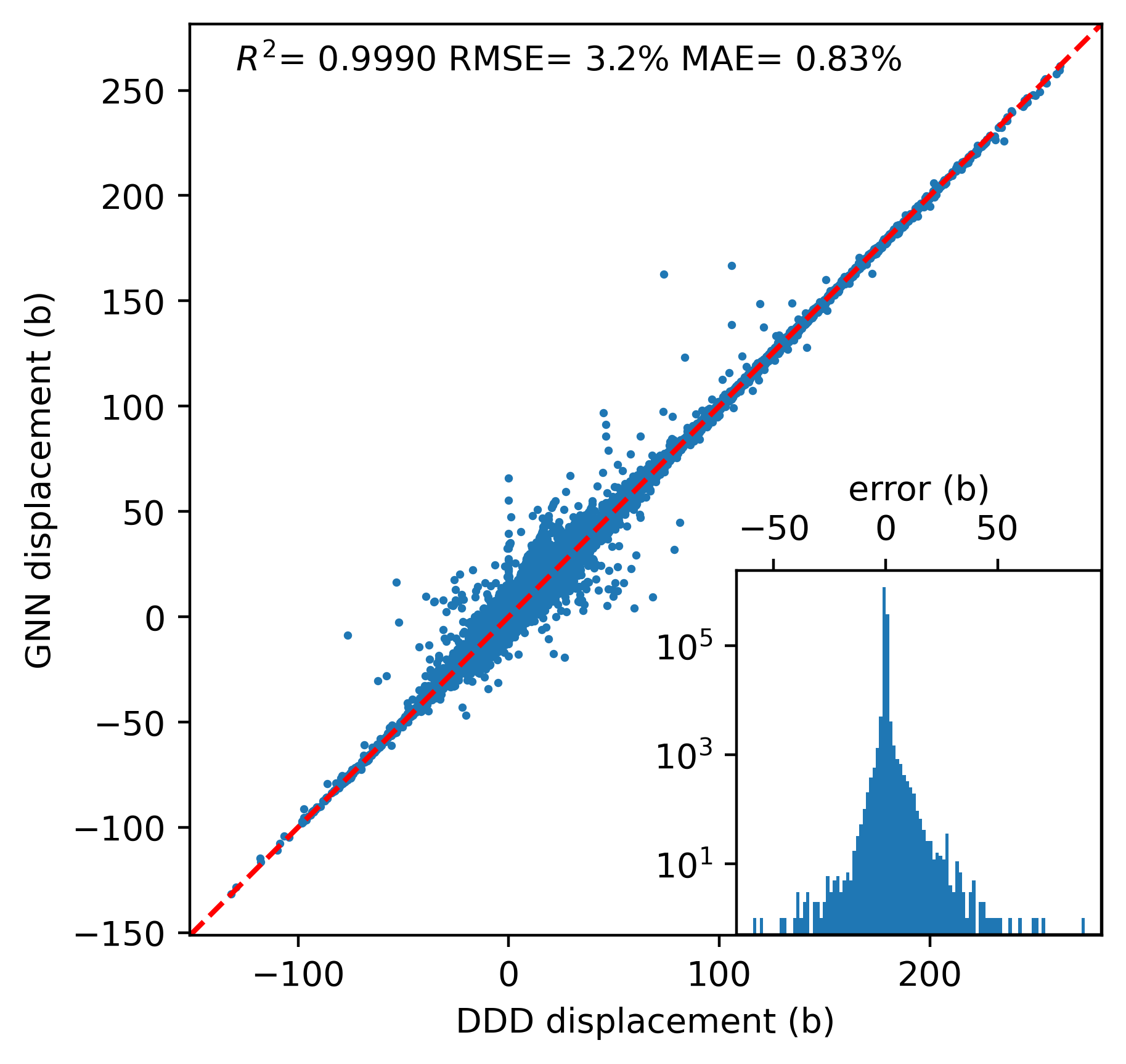}
\end{center}
  
  %\vspace*{-0.5cm}
  \caption{GNN training results. (a) Validation loss vs.\ training iterations (epochs) at different number of message-passing layers $K$. (b) Predicted versus ground-truth components of the nodal displacements in unit of $b$ in the validation set; Inset: histogram of validation error.}
\label{fig:training}
\end{figure}

\subsection{GNN architecture and training}

The GNN model was trained within the PyTorch framework.
The DDD dataset was split on a 90:10 ratio between a training and a validation set. The latter was used to monitor the validation loss during training. The following data pre-processing procedure was adopted: the input applied and long-range stresses were normalized by a factor of $10$ MPa, node positions were normalized by a factor of $100b$, and the target properties (time-integrated displacements) were centered to their mean and normalized over their standard deviation before being fed into the GNN, and then scaled back to their actual values, both using statistics determined from the training dataset.

A grid search was performed on the most important hyper-parameter $K$, the number of message passing layers, between $1 \leq K \leq 4$. As shown in Fig.~\ref{fig:training}a, $K=3$ was found to yield a good regression accuracy, while larger $K=4$ did not to lead to meaningful improvement.
The latent space dimension was similarly chosen to be 96 while larger latent spaces added to more computational costs but no noticeable accuracy. The model was trained as a regression task against DDD integrated node displacements using the mean-square error loss function and the Adam optimizer with weight decay of $5 \times 10^{-3}$ \cite{Adam, Adamw}. Training was performed with an initial learning rate of $3 \times 10^{-4}$, which was reduced by 70\% upon validation loss plateaus until a minimum value of $10^{-6}$, for 12 hours with batch size of 8 on a single NVidia V100 GPU.

Fig.~\ref{fig:training}b compares the predicted vs ground-truth displacements of the validation dataset.
Good accuracy is seen from the RMSE of $0.64b$ (3.2\% relative to the standard deviation of the training set) and MAE of $0.17 b$ (0.83\%). Still, there exists a small fraction of outliers (1.7\% with absolute error above $1b$), although the scatter plot in Fig.~\ref{fig:training}b and the log scale in the inset exaggerate such rare outliers.
Such outliers are common in traditional and ML interatomic potential fittings but do not necessarily translate into large errors in subsequent MD simulations \cite{Behler2021}. From the results presented in the next section, we also find that these outliers do not lead to deteriorated accuracy of GNN predictions in practice.

\section{Results} \label{sec:results}

\begin{figure}[t]
  \begin{center}
    \includegraphics[width=0.48\textwidth]{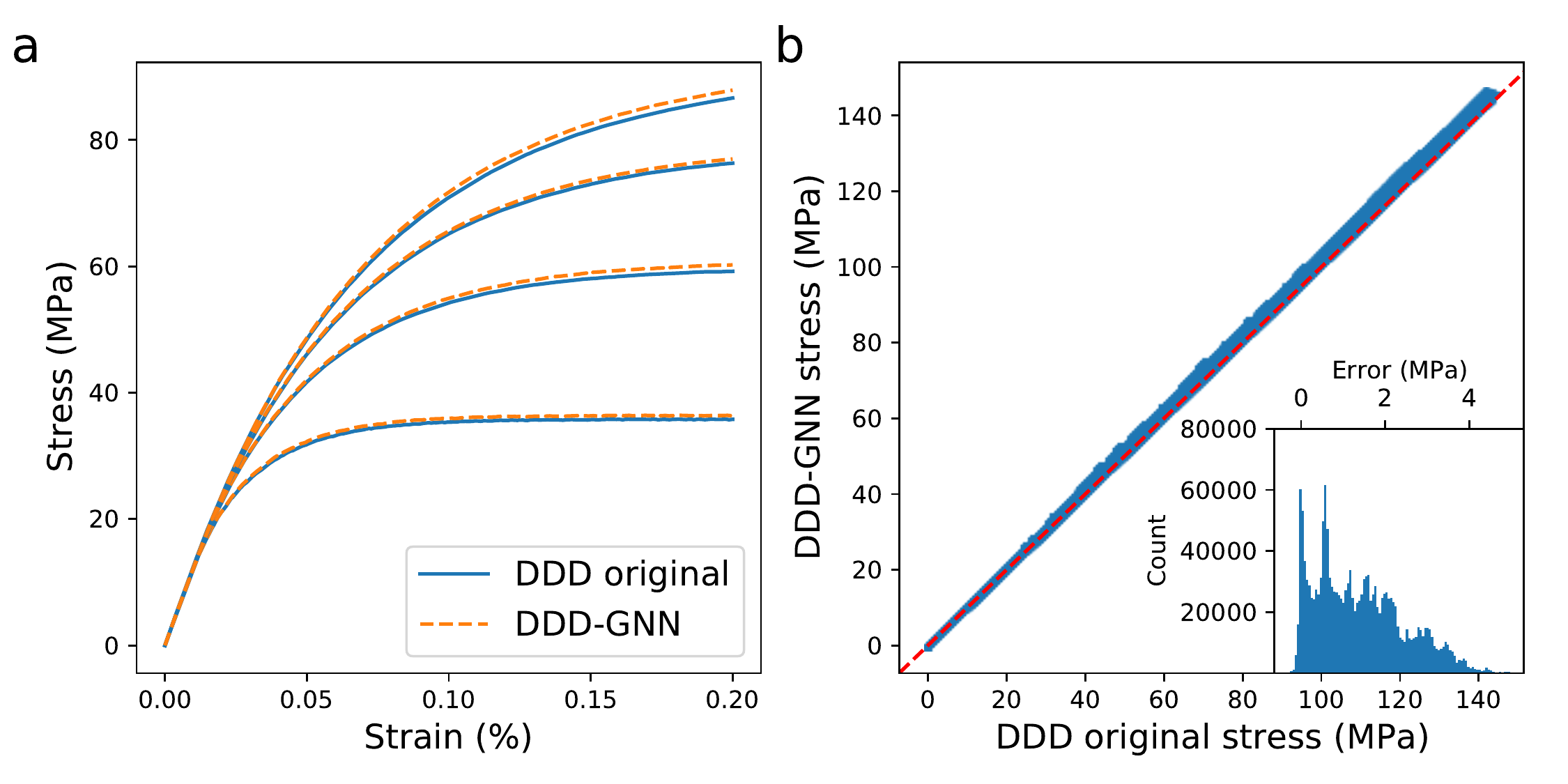}
  \end{center}
  %\vspace*{-0.5cm}
  \caption{DDD-GNN test results for simulations in the drag-controlled regime. (a) Example of stress-strain curves obtained with ground-truth DDD simulation and as predicted by the DDD-GNN approach using the same simulation parameters. (b) Comparison of stress values along the full simulation trajectories between unseen ground-truth DDD simulations and DDD-GNN predictions; Inset: histogram of the stress error predictions of the DDD-GNN model.}
\label{fig:valid_drag}
\end{figure}

In this section we present results of our trained DDD-GNN model. To validate our model, we first generated a test dataset composed of a series of 100 new, unseen ground-truth DDD simulations using the same range of obstacles densities and strain rates than used in the training process (i.e. in the drag-controlled regime), but using different seeds for generating the random obstacles arrays, obstacles densities, and strain rate values.
The exact same seeds were then used to repeat the simulations using the DDD-GNN model.
Each simulation was run until reaching a strain of $0.2\%$, i.e. for trajectories of 5,000 to 50,000 time-steps depending on the straining rate.

Results for the direct comparison of the ground-truth DDD and DDD-GNN simulations are shown in Fig.~\ref{fig:valid_drag}. An example of the stress/strain curves computed by the DDD and DDD-GNN methods are presented in Fig.~\ref{fig:valid_drag}a. The curves are following each other very closely, indicating the quality of the DDD-GNN predictions.
To better examine the accuracy of DDD-GNN, Fig.~\ref{fig:valid_drag}b compares the ground-truth DDD stress values along the full trajectories for the 100 new simulations to the values predicted by the DDD-GNN. The agreement is remarkable, with MAE of 1.4 MPa and RMSE of 1.75 MPa. The histogram of stress errors shown in the inset indicates that all stress predictions of the DDD-GNN model fall within a $\sim 5$ MPa error interval.

\begin{figure}[t]
  \begin{center}
    \includegraphics[width=0.48\textwidth]{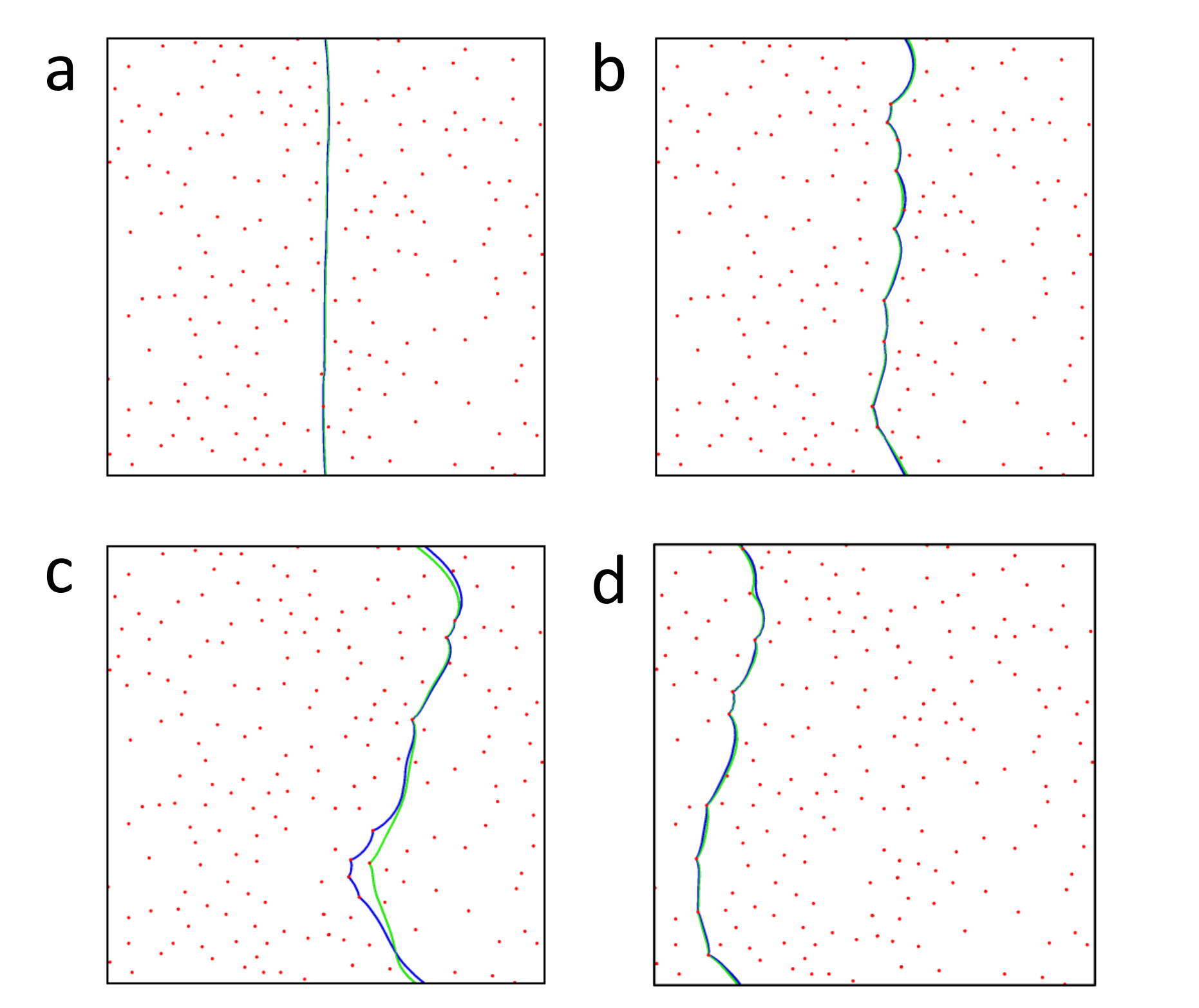}
  \end{center}
  %\vspace*{-0.5cm}
  \caption{Example of the comparison of a ground-truth DDD trajectory (green line) and the corresponding DDD-GNN predicted trajectory (blue line) at different instants in time. Obstacles are marked as red dots. The DDD-GNN mobile dislocation is seen to closely follow the ground-truth trajectory as it glide through the \rv{array} of obstacles and across periodic boundaries.}
\label{fig:valid_traj}
\end{figure}

To illustrate the accuracy of the DDD-GNN model at the microstructural level, Fig.~\ref{fig:valid_traj} shows an example of the dislocation line trajectory predicted by DDD-GNN (blue line) and compared to the ground-truth DDD simulation (green line) under the same conditions. The different snapshots show that the dislocation lines closely follow each other as they glide through the \rv{array} of obstacles.

To further assess the capabilities of the DDD-GNN approach, we now perform 10 additional new simulations in a different parameter space. Specifically, we reduce the size of the simulation box along the glide plane normal direction by a factor 5 to a value of $H = 0.4 {\rm \mu m}$. This change -- which corresponds to increasing the density of the mobile dislocation -- has the effect of reducing the drag stress $\tau^{\rm drag}$ to values comprised between 3.8 and 38 MPa, i.e. below or on par with the obstacle stress values. Thus, these additional simulations are now probing the different regime of obstacle-controlled glide, where obstacles are expected to be the controlling factor in defining the flow-stress of the crystals. To further test the model these simulations were run for longer trajectories of 100,000 time-steps.

\begin{figure}[t]
  \begin{center}
    \includegraphics[width=0.48\textwidth]{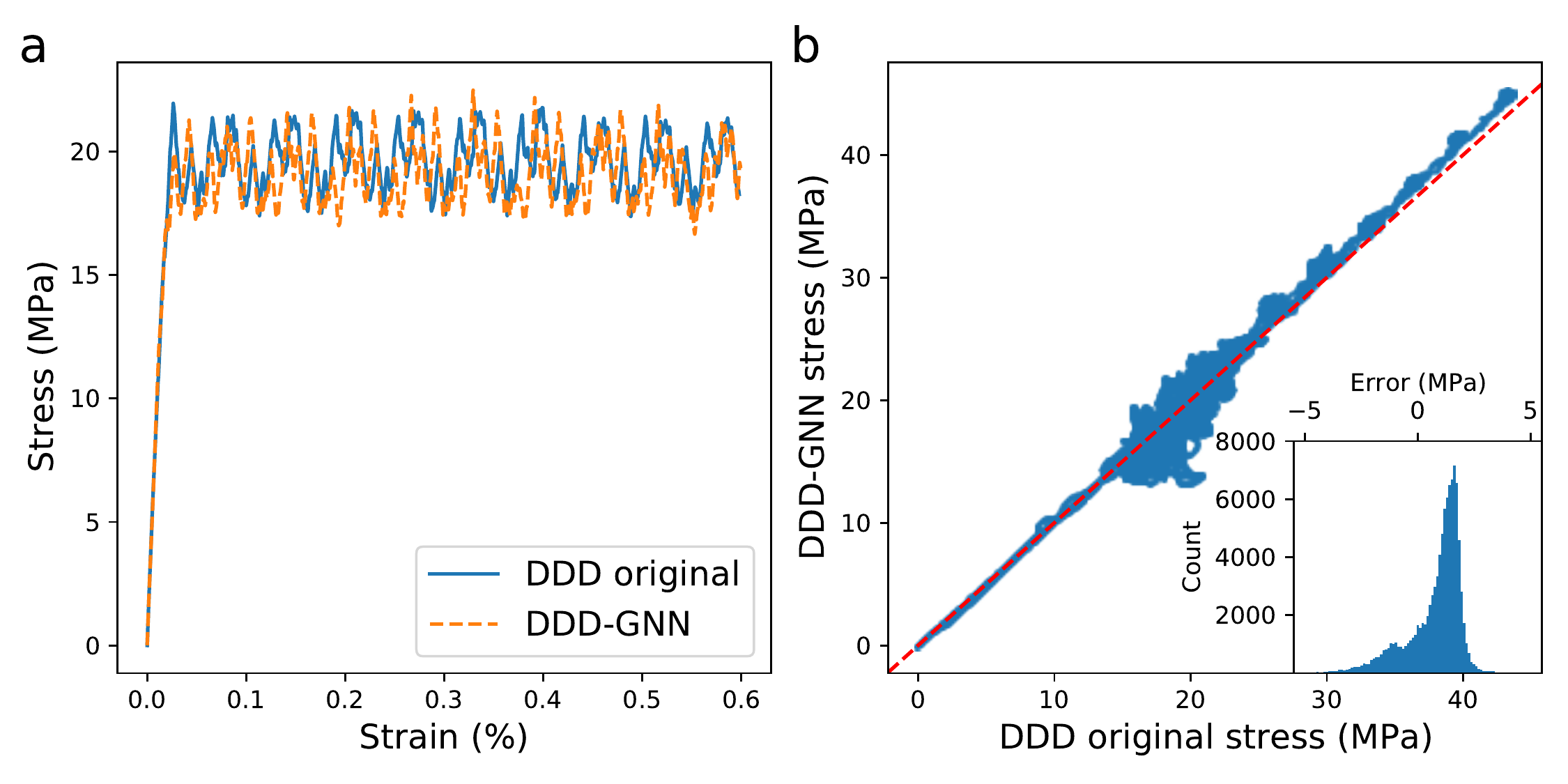}
  \end{center}
  %\vspace*{-0.5cm}
  \caption{DDD-GNN test results for simulations in the obstacle-controlled regime. (a) Example of a stress-strain curve obtained with ground-truth DDD simulation and as predicted by the DDD-GNN approach for 100,000 time-steps using the same simulation parameters. (b) Comparison of stress values along the full simulation trajectories between unseen ground-truth DDD simulations and DDD-GNN predictions; Inset: histogram of the stress error predictions of the DDD-GNN model.}
\label{fig:valid_obs}
\end{figure}

Results for the DDD-GNN predictions in this regime are shown in Fig.~\ref{fig:valid_obs}. An example of the stress-strain curves obtained for this set of obstacles-controlled simulations is shown in Fig.~\ref{fig:valid_obs}a. As expected, the stress response now exhibits relatively large fluctuations compared to the smooth responses obtained in the drag-controlled regime, e.g. Fig.~\ref{fig:valid_drag}a. Interestingly, the DDD-GNN is seen to still capture very well the macroscopic behavior in this regime. While both DDD and DDD-GNN stress predictions do not strictly overlap anymore, the average flow-stress values and magnitude of the fluctuations are similar.
For example, for the simulation presented in Fig.~\ref{fig:valid_drag}a the average DDD stress in the flow regime is of $19.58 \pm 1.09$ MPa while that predicted by the DDD-GNN model is of $19.13 \pm 1.19$ MPa.
Generally, we find that the differences between the ground-truth DDD simulations and the DDD-GNN predictions in terms of their fluctuations are no larger than are differences between two ground-truth simulations with same parameters but different seeds. That is, the DDD-GNN model is statistically equivalent to the ground-truth DDD.

Stress error predictions for the whole trajectories of the 10 obstacles-controlled simulations are shown in Fig.~\ref{fig:valid_obs}b. The agreement between the trained model and the ground-truth DDD is very good overall, with MAE of 1.28 MPa and RMSE of 1.46 MPa. The wider cloud of points in the stress region near 20 MPa corresponds to fluctuations of the kind reported in Fig.~\ref{fig:valid_obs}a. For this choice of the parameters, the amplitude of these fluctuations is maximum for stresses of $\sim 20$ MPa, and decreases with increasing stresses. The region of lower stresses mostly corresponds to the elastic regime in which no significant dislocation motion is taking place. The fact that an excellent agreement is observed here additionally indicates that the DDD-GNN model correctly predicts the initial activation of dislocation motion (onset of plasticity).

\section{Discussion} \label{sec:discussion}

The results of the DDD-GNN framework shown in the previous section demonstrate the viability and potential of such a ML approach for DDD simulations. 
\rv{
Results obtained for the obstacle-controlled regime are interesting because the GNN has not been explicitly trained on examples in this regime. 
The fact that the model is still able to provide accurate predictions suggests that the GNN-based integrator has learned the correct physics during the training process (no overfitting) and is thus able to "generalize" well under this different regime.
Here, we note however that the model is not extrapolating from the training examples in the drag-controlled regime. The obstacle-controlled regime is generally characterized by a lower driving stress for the dislocations compared to the drag-controlled regime. Since the training trajectories (in the drag-controlled regime) contain snapshots of simulations loaded at various strain rates from an initial zero stress state, examples of dislocation behaviors at lower stresses (e.g. during the elastic regime) were implicitly included in the training data, and thus learned during training. As such the model is interpolating in both the regimes considered here, hence the quality of the predictions.
Conversely, we do not expect that the model would generalize as well in higher-stress regimes -- e.g. as would be obtained in simulations with higher obstacles densities or smaller values of the breakaway angle -- for which no example of dislocation behaviors was present in our training dataset. Therefore, as with any ML approach, applying the model to remote regimes would generally require to retrain the model with new datasets that contain examples of the behavior in these different regimes.}

One of the most important aspects of the model is its ability to correctly capture the long-term dynamics of the dislocation system, e.g. to accurately predict the \rv{evolution (and saturation) of the} flow stress.
This is significant because it shows that as proposed, the DDD-GNN model seems not to suffer from stability issues typically encountered in ML models \cite{arxiv.2108.13624, wang2021long}.
In many ML applications, accumulation of errors from individual time-step predictions often leads to large divergences with respect to the ground-truth simulations over long-term trajectories.
Here we did not observe any sign of such divergence over the course of up to 100,000 consecutive time-steps simulated in our test set.
The stability of our model may be imparted by our hybrid approach, in which only the time-integration stage of the simulation is accelerated with the GNN model, while the rest of the simulation cycle is performed using the native implementation.
It is thus possible that our hybrid scheme, e.g. in which topological operations are still treated explicitly following local rules \cite{Arsenlis07}, acts as a stabilizer for the simulation dynamics.

%The other advantage of the hybrid approach is that it allows us to circumvent the difficult issue of dealing with topological changes directly with GNN. While some approaches have been proposed to handle such events \cite{}  (not even sure this really exists, does it?), it would significantly complexify our approach.
%Alternatively, taking advantage of the modular nature of DDD allows us to greatly simplify the problem and first focus on the relevant stages of the simulation that can greatly benefit from a ML approach.

Our hybrid approach takes advantage of the modular nature of DDD, reusing the well-developed techniques of dealing with topological changes in DDD as well as highly-optimized long-range interaction computations, while focusing on improving the simulation efficiency by replacing the computationally most expensive operations, i.e.\ short-range interactions and time integration, with a machine-learned surrogate model.
We therefore envision that the primary benefit of using this approach would lie in the acceleration of the simulations. In DDD, it was recently shown that efficient time-integration algorithms for dislocation motion is key to overcome the long-standing time-scale limitations of the approach \cite{Sills14, Sills16}. Here, instead of developing more sophisticated mathematical time-integration procedures, e.g. such as implicit integrators \cite{gardner2015implicit, peterffy2020efficient} or the subcyling approach \cite{Sills16}, we take a different route and propose instead to rely on a learnable GNN model trained to predict nodal displacements given the current state of the dislocation network. For instance, in this work our model was trained using a multi-stepping time-integrator. Conceptually, our model can thus be regarded as an approach to turn a multi-stepping explicit integrator into an expedient, data-driven implicit time-integrator.

%And while our model was trained using a multi-stepping time-integrator in this work, it could in principle be trained to advance nodal positions as computed with any ground-truth time-integrator of arbitrary degree of sophistication.

In our simulations we find that our GNN approach allows to accelerate the time-integration procedure by a factor $\sim 10$ compared to the ground-truth multi-stepping integrator on a single CPU core.
In addition, our approach is particularly well suited for execution on the new generation of heterogeneous computing platforms with hardware accelerators (e.g. GPU), to shift the bulk of the floating-point computational burdens to neural network subroutines that are heavily optimized for such hardware. And in contrast to traditional integrators, GPU execution of the model is readily available in existing GNN frameworks, and thus does not require any separate, dedicated implementation \cite{bertin2019gpu}.
We also expect that substantial additional computational savings are to be made in more complex scenarios, e.g. in the case of bulk strain-hardening simulations where short-range interactions are more numerous and fully dominate the computation load.
In this first study, we chose to limit ourselves and apply the model to \rv{dislocation-obstacle glide} simulations to demonstrate the potential of the approach. While the framework presented in Section \ref{sec:DDDGNN} is fully general, in practice applying it to bulk DDD simulations is slightly more challenging and requires some additional development. This is out of the scope of this paper and will be the subject of a dedicated future publication.

Finally, we point out that the DDD-GNN framework is general is the sense that it is agnostic to the precise physics of dislocation motion present in the training data. Specifically, as proposed the model is devised to bypass both the explicit calculation of short-range interaction forces and nodal velocities. For instance, this means that the model requires no assumption nor knowledge about the underlying form and level of complexity of the mobility function $\mathcal{M}$, see Eq.~\eqref{eq:EOM}.
An interesting perspective is then to employ the DDD-GNN framework to learn dislocation motion directly from MD trajectories. A major advantage would be that the model could be trained to realistically approximate dislocation physics without the need to posit functional forms that constrain the mobility operator (e.g. which may be biased with a detrimental impact on the DDD predictions).
While many challenges would need to be solved in order to achieve this in practice, the DDD-GNN  framework nonetheless opens the way to promising future applications, e.g. to include more complex, accurate dislocation physics within the DDD framework.

\section{Conclusion} \label{sec:conclusion}

We introduced a novel DDD-GNN framework aimed at bypassing the time-consuming, explicit nodal force/velocity calculations in dislocation dynamics simulations by directly predicting nodal displacements instead.
To achieve this, our framework relies on two main ideas.
First, we propose an hybrid approach whereby a GNN model is used to fully replace the time-integration procedure, while the rest of the DDD cycle, e.g. involving topological operations, is executed following the traditional DDD workflow.
Second, we take advantage of the short/long range partitioning of elastic interactions traditionally used in DDD approaches to build our model. The idea is that short-range interactions, which are most computationally expensive, are fully defined by the local dislocation line configuration and can thus be learned accurately via the message-passing algorithm, while longer-range interactions, which are relatively inexpensive to compute in comparison, can be provided as segments attributes of the dislocation configuration.

To showcase the capability of our approach, we applied it to the case of a mobile dislocation gliding through a random \rv{array} of obstacles. To train our model, we performed a series of ground-truth DDD simulations with varying obstacles densities and imposed straining rates. During the simulated trajectories we recorded instantaneous configurations and nodal displacements at frequent interval in time to generate the training data for the model. Our trained model was then implemented in the DDD code ParaDiS and results were compared to ground-truth simulations. We found an excellent agreement between the DDD-GNN predictions and unseen ground-truth DDD simulations, and showed that our model is stable and accurately captures the long-term dynamics of the simulated dislocation systems, even when exercised in a regime different from that used in the training process.

We further showed that the novel framework proposed in this work has the potential to significantly accelerate DDD simulations. The approach outlined in this work paves the way for addressing the more complex case of bulk strain-hardening DDD simulations, and opens new avenues for incorporating more complex physics that does not require human intuition.

\section*{Acknowledgement}
The authors would like to thank Dr. Cheol Woo Park for fruitful discussions.
NB acknowledges support by the Laboratory Directed Research and Development (LDRD) program (22-ERD-016) at Lawrence Livermore National Laboratory (LLNL). FZ was supported by the Critical Materials Institute, an Energy Innovation Hub funded by the U.S. Department of Energy, Office of Energy Efficiency and Renewable Energy, and Advanced Manufacturing Office.
Computing support for this work came from LLNL Institutional Computing Grand Challenge program.
This work was performed under the auspices of the U.S. Department of Energy by LLNL under contract DE-AC52-07NA27344.

\appendix

%\section{Section}

%\section*{References}
\bibliography{ref}

\end{document}